\documentstyle[prl,multicol,aps,epsf]{revtex}

\begin{document}

\epsfverbosetrue

\title{Stable embedded solitons}

\author{Jianke Yang}
\address{Department of Mathematics and Statistics, 
University of Vermont, Burlington, VT 05401, USA}
\maketitle

\begin{abstract}

{\em Stable} embedded solitons are discovered
in the generalized third-order nonlinear Schr\"odinger equation. 
When this equation can be reduced to 
a perturbed complex modified KdV equation, we developed
a soliton perturbation theory which shows that a continuous family of 
sech-shaped embedded solitons exist and are nonlinearly 
{\em stable}. These analytical results are confirmed by our
numerical simulations. 
These results establish that, contrary to previous beliefs, 
embedded solitons can be robust despite
being in resonance with the linear spectrum. 
[PACS: 42.65.Tg, 05.45.Yv.]
\end{abstract}

\begin{multicols}{2}
\narrowtext

Pulse propagation in optical fibers attracts a lot of 
attention these days. Pico-second pulses are well described by
the nonlinear Schr\"odinger (NLS) equation which accounts for 
the second-order dispersion and self-phase modulation. 
But for femtosecond pulses, other physical effects such as the third-order
dispersion and self-steepening become non-negligible. The physical model which
incorporates these additional effects is \cite{agrawal,hasegawa}
\small
\begin{equation} \label{psi}
i\left[ \psi_z+\beta |\psi|^2\psi_\tau+\mu (|\psi|^2)_\tau \psi \right]
+\frac{1}{2}\psi_{\tau\tau}+|\psi|^2\psi+i\psi_{\tau\tau\tau}=0. 
\end{equation}
\normalsize
Here $\psi$ is the envelope of the electric field, 
$z$ is the distance, $\tau$ is the retarded time, 
$\beta$ and $\mu$ are the self-steepening coefficients \cite{agrawal}. 
All quantities have been normalized. 
For optical pulses, $\beta=\mu$ \cite{agrawal,hasegawa}. 
In this paper, we allow $\beta$ and $\mu$ to be different 
for the sake of mathematical analysis. 
The Raman effect, which is dissipative in nature, is 
also non-negligible \cite{agrawal,hasegawa}. 
It is not included in the model (\ref{psi}) 
because we want to focus our attention on the other physical effects 
in this paper. 

The third-order dispersion term in Eq. (\ref{psi}) is significant because
it qualitatively changes the linear dispersion relation of Eq. (\ref{psi}). 
Its effect on the NLS soliton is to generate
continuous-wave radiation and causes the soliton to decay
\cite{agrawal,hasegawa,wai90,grimshaw,boydbook}. 
However, solitary waves of Eq. (\ref{psi}) which are embedded 
inside the linear spectrum do exist in certain parameter regimes
\cite{klauder,akylas,gromov0,yang_akylas}, and
such waves are called embedded solitons \cite{YangPRL99,champneys}. 
To see why a solitary wave of Eq. (\ref{psi}) has to be 
an embedded soliton, we substitute solitary waves of the form
$\Psi(\tau-Vz)e^{i\lambda z}$ into (\ref{psi}), 
where velocity $V$ and frequency $\lambda$ are constants. 
We readily find
that for any frequency $\lambda$, the linear equation for
$\Psi$ allows oscillatory solutions. Thus, all $\lambda$ lies in
the continuous spectrum of Eq. (\ref{psi}), hence the solitary wave
must be an embedded soliton. 
Stability of these embedded solitons is clearly an important issue. 
Previous analytical studies have shown 
that if embedded solitons are isolated in a conservative system, 
they are at most semi-stable, i.e., the perturbed soliton
persists or decays depending on whether the initial energy is higher or lower
than that of the embedded soliton \cite{YangPRL99,YangStudies01,PeliYang}. 
A physical explanation for it is as follows \cite{YangPRL99}. 
If the initial state has energy higher than the embedded soliton, 
it just sheds extra energy through tail radiation and asymptotically approaches
this embedded soliton; but if the initial state has lower energy than
the embedded soliton, the energy loss (through radiation) drives the solution
away from the embedded soliton, resulting in instability. 
However, embedded solitons in Eq. (\ref{psi}) are {\em not isolated}: 
they exist as a continuous family, parameterized by their 
energy \cite{gromov0,yang_akylas}. Thus, the above physical 
argument for semi-stability does not apply. 
Numerical results suggest that 
the families of double-hump embedded solitons in Eq. (\ref{psi})
are still semi-stable \cite{yang_akylas}, but 
the family of single-hump embedded solitons are fully stable \cite{gromov}.

In this paper, we show that 
single-hump embedded solitons in Eq. (\ref{psi}) are fully stable
when $|\beta-6|\ll 1$ and $|\mu|\ll 1$. To our knowledge, this
is the first time {\em stable} embedded solitons
are rigorously established in the literature. This result indicates that 
embedded solitons can be as robust as conventional (non-embedded) solitons, 
contrary to the previous belief. The method we will use is to 
develop a soliton perturbation theory for the (integrable) complex 
modified KdV (CMKdV) equation, supplemented by numerical simulations. 

We first employ the variable transformation 
\begin{equation}
\psi=e^{\frac{1}{6}i(\tau-\frac{1}{18}z)} \sqrt{\frac{6}{\beta}}
\: u(z,\xi), \;\;\; \xi\equiv \tau-\frac{1}{12}z, 
\end{equation}
upon which Eq. (\ref{psi}) becomes 
\begin{equation} \label{uold}
u_z+u_{\xi\xi\xi}+6|u|^2u_\xi=i \alpha |u|^2u-\gamma (|u|^2)_\xi u,
\end{equation}
where $\alpha=(6-\beta)/\beta$, and $\gamma=6\mu/\beta$. 
If $\alpha=\gamma=0$, the above equation is the CMKdV equation
which is completely integrable by the inverse scattering method 
\cite{hirota}. It admits sech-shaped soliton solutions whose amplitudes
and velocities are free parameters (see below). In this paper, we
consider the case $|\alpha|, |\gamma| \ll 1$, i.e., $|\beta-6|\ll 1$ and $|\mu|\ll 1$. 
In this limit, soliton evolution in the
perturbed CMKdV equation (\ref{uold}) can be studied by 
a soliton perturbation theory. For this purpose, we denote
$\alpha=\epsilon \alpha_0, \gamma=\epsilon \gamma_0$, where $|\epsilon| \ll 1$. 
When $\epsilon=0$, Eq. (\ref{uold}) has the following soliton solutions
\begin{equation} \label{u0}
u_0(z,\xi)=U_0(\theta)e^{i\lambda z}, \;\; \theta\equiv \xi-Vz, 
\end{equation}
where 
\small
\begin{equation}\label{U0}
U_0(\theta)=r \mbox{sech} r\theta e^{ik\theta}, \;\; V=r^2-3k^2, \lambda=-2k(r^2+k^2).
\end{equation}
\normalsize
Here amplitude $r$ and frequency $k$ are free parameters. 
When $0\ne |\epsilon| \ll 1$, these solitons will deform due to perturbations
on the right hand side of Eq. (\ref{uold}), and their amplitudes and frequencies
will undergo slow evolution with respect to $z$. 
Below, we use the soliton perturbation theory to derive this evolution. 

First, we write the solution in the form 
\begin{equation} \label{uU}
u(z,\xi)=e^{i\int_0^z \lambda dz} U(z,\theta), \;\; \theta=\xi-\int_0^z V dz. 
\end{equation}
Substituting this form into (\ref{uold}), we find the equation for $U(z, \theta)$ as
\begin{equation} \label{U}
U_z+i\lambda U-VU_\theta+U_{\theta\theta\theta}+6|U|^2U_\theta=\epsilon F(U), 
\end{equation}
where 
\begin{equation} \label{F}
F(U)=\left[i\alpha_0|U|^2-\gamma_0(|U|^2)_\theta\right] U.
\end{equation}
Next, we expand the solution $U$ into a perturbation series
\begin{equation} \label{expansion}
U(z,\theta)=U_0(\theta)+\epsilon U_1(z,\theta)+\epsilon^2 U_2(z,\theta) +\dots,
\end{equation}
where $U_0$ is given in (\ref{U0}). 
When this series is substituted into Eq. (\ref{U}), at order 1, the 
equation is satisfied automatically. At order $\epsilon$, we obtain
the linear inhomogeneous equation for $U_1$ as
\begin{equation} \label{U1}
\left(\frac{\partial}{\partial z}+{\cal L}\right)
\left(\begin{array}{c} U_1 \\ U_1^* \end{array}\right)
= \left(\begin{array}{c} F(U_0) \\ F^*(U_0) \end{array}\right), 
\end{equation}
where ${\cal L}$ is the linearization operator of Eq. (\ref{U}), 
and superscript ``*'' denotes the complex conjugation. 
At initial distance $z=0$, $U_1=0$; when $z\gg 1$, 
$U_1$ approaches a steady state with continuous-wave tails at infinity. 
Generation of continuous-wave tails is a distinctive feature of 
embedded solitons under perturbations \cite{YangPRL99}. 
These tails have the form $he^{-2ik\theta}$, where $h$ is the tail
amplitude. Due to the Sommerfeld radiation condition, these tails
can only appear at $\theta \ll -1$, not at $\theta \gg 1$. In other words, 
\begin{equation} \label{U1limit}
\lim_{z \to \infty,\; \theta \ll -1}U_1=he^{-2ik\theta}, 
\;\; \lim_{z \to \infty,\; \theta \gg 1}U_1=0.
\end{equation}
One of the key steps in the soliton perturbation theory is to determine
the amplitude $h$ of continuous-wave tails in $U_1$. This can be done
by solving Eq. (\ref{U1}) directly, starting with the zero initial condition
\cite{PeliYang}. 
The way to do it is to expand the solution into the complete set of eigenfunctions
for the linearization operator ${\cal L}$, which are just the squared 
eigenstates of the Zakharov-Shabat system with the soliton potential
\cite{YangJMP}. This we have done. 
But a much simpler way to derive $h$ 
is to just consider the steady-state solution $U_1$. This allows us to 
drop the $z$ derivative in Eq. (\ref{U1}), and use only solvability conditions
to obtain $h$. This idea has been used before \cite{YangStudies01}. 

To pursue this latter approach, we need the bounded eigenfunctions 
of the adjoint linearization operator ${\cal L}^A$ with zero eigenvalue. 
Here the inner product used to define an adjoint operator is 
\begin{equation}
\langle f(\theta), g(\theta)\rangle = \int_{-\infty}^{\infty} f(\theta)^T g(\theta)d\theta,
\end{equation}
where the superscript ``$T$'' represents the transpose of a vector. 
Under this inner product, the adjoint operator ${\cal L}^A$ can be readily obtained. 
Eigenfunctions of ${\cal L}^A$ are simply a different set of squared 
Zakharov-Shabat eigenstates \cite{YangJMP}. With this in mind, we can 
easily show that ${\cal L}^A$ has four bounded eigenfunctions for
zero eigenvalue --- two localized and two non-localized. 
The localized (discrete) eigenfunctions are
\begin{equation} \label{eigen1}
\Phi_{1}=\left[\begin{array}{c} U_{0\theta}^* \\ -U_{0\theta}\end{array}\right], 
\;\; 
\Phi_{2}=\left[\begin{array}{c} U_{0}^* \\ U_{0}\end{array}\right].
\end{equation}
They are associated with the translational and phase invariances 
of solitons in the CMKdV equation. 
The non-localized (continuous) eigenfunctions are 
\begin{equation} 
\Phi_{3}=\left[\begin{array}{c} r^2\mbox{sech}^2r\theta e^{-4ik\theta} \\
(r\tanh r\theta+3ik)^2e^{-2ik\theta}\end{array}\right], 
\end{equation}
\begin{equation} 
\Phi_{4}=\left[\begin{array}{c} (r\tanh r\theta-3ik)^2e^{2ik\theta} \\
r^2\mbox{sech}^2r\theta e^{4ik\theta}\end{array}\right].
\end{equation}
Now we take the inner product between 
Eq. (\ref{U1}) (with no $z$ derivative) and each of the above four eigenfunctions. 
In view of the asymptotics (\ref{U1limit}), we see that 
the first two inner-product equations are simply
\begin{equation}\label{cond12}
\left\langle \Phi_{1}, 
\left(\begin{array}{c} F(U_0) \\ F^*(U_0) \end{array}\right) \right\rangle
=\left\langle \Phi_{2}, 
\left(\begin{array}{c} F(U_0) \\ F^*(U_0) \end{array}\right) \right\rangle=0,
\end{equation}
which are satisfied automatically when the form (\ref{F}) for $F(U)$ is
utilized. The last two inner-product equations are equivalent. 
After simple calculations, these equations 
give a formula for the amplitude $h$ as
\begin{equation}\label{h}
h=-\frac{1}{4}i\pi(\alpha_0+2\gamma_0 k)\frac{r-3ik}{r+3ik}\mbox{sech} \frac{3\pi k}{2r}.
\end{equation}

When the tail amplitude $h$ has been derived, we can proceed to order $\epsilon^2$. 
Again, by applying the solvability conditions for solution $U_2$, 
the dynamical equations for soliton parameters $r$ and $k$ 
will be obtained (see \cite{YangStudies01} for an example). 
An alternative way to derive these dynamical equations, 
which is simpler and physically more insightful, is by using conservation laws
\cite{YangStudies01}. In this latter way we will proceed. 

Eq. (\ref{uold}) admits the following two conservation laws: 
\small
\begin{equation} \label{I1}
I_1=\int_{-\infty}^{\infty} |u|^2 d\xi, 
\end{equation}

\vspace{-.3cm}
\scriptsize
\begin{equation} \label{I2}
I_2=\int_{-\infty}^{\infty} \left\{\frac{1}{2}i\alpha (u_\xi u^*-uu_\xi^*)
-\gamma|u_\xi|^2+\gamma(1+\frac{1}{3}\gamma)|u|^4\right\} d\xi. 
\end{equation}
\normalsize
When Eq. (\ref{uU}) and the perturbation series (\ref{expansion})
are substituted into these conservation laws and 
terms up to order $\epsilon^2$ retained, the first conservation law
becomes 
\scriptsize
\begin{equation} \label{I1b}
\frac{d}{dz}\int_{-\infty}^{\infty} \left\{
|U_0|^2 + \epsilon (U_0U_1^*+U_0^*U_1)
+\epsilon^2 (U_0U_2^*+U_0^*U_2) + \epsilon^2 |U_1|^2\right\}=0. 
\end{equation}
\normalsize
To proceed further, the following facts are noted: 
(i) $|U_0|^2$ changes slowly (on the distance scale $\epsilon^2 z$);
(ii) the $U_1$ field is driven by the inhomogeneous term in Eq. (\ref{U1})
    and quickly becomes stationary in the soliton ($U_0$) region; the $U_2$
    field is similar; 
(iii) the $U_1$ field develops a continuous-wave tail of amplitude $h$ 
     on the left-hand-side
     of the soliton [see Eq. (\ref{U1limit})]; this tail travels at its 
     group velocity $v_g=-12k^2$; in the moving coordinate $\xi$, its relative
     velocity is $v_g-V=-(r^2+9k^2)$ [see Eq. (\ref{U0})]. 
The fact (ii) means that the $z$ derivatives of the integrals of
the second and third terms in Eq. (\ref{I1b}) are zero and can be dropped. 
The fact (iii) means that the $z$ derivative of the integral of 
the last term in (\ref{I1b}) is $\epsilon^2(r^2+9k^2)|h|^2$. 
When these results
and the $U_0$ formula (\ref{U0}) are utilized, the first conservation law
(\ref{I1b}) gives the dynamical equation for the soliton amplitude $r$. 
Similar calculations for the second conservation law produces the
dynamical equation for the soliton frequency $k$. 
When the $h$ formula (\ref{h}) is substituted, and 
($\epsilon \alpha_0, \epsilon \gamma_0$) replaced by the
physical parameters ($\alpha, \gamma$), 
the dynamical equations for $r$ and $k$ finally take the form
\small
\begin{equation} \label{r}
\frac{dr}{dz}=-\frac{1}{32}\pi^2 (\alpha+2\gamma k)^2 (r^2+9k^2) \mbox{sech}^2\frac{3\pi k}{2r}, 
\end{equation}
\begin{equation} \label{k}
\frac{dk}{dz}=\frac{1}{32r}\pi^2 (\alpha+2\gamma k) [3\alpha k-\gamma (r^2+3k^2)]
 (r^2+9k^2) \mbox{sech}^2\frac{3\pi k}{2r}, 
\end{equation}
\normalsize
and they govern the soliton evolution under small perturbations $|\alpha|, |\gamma| \ll 1$
in Eq. (\ref{uold}). 
These two equations are the main analytical results of this paper. 

Dynamical equations (\ref{r}) and (\ref{k}) admit a continuous family
of fixed points: 
\begin{equation} \label{fixpt}
k=-\frac{\alpha}{2\gamma}, \;\;\; r\; \mbox{is arbitrary}. 
\end{equation}
These fixed points correspond to a continuous family of 
embedded solitons (\ref{u0}) in the original wave equation (\ref{uold}). 
Their amplitudes
$r$ are arbitrary, while their speeds $V$ depend on $r$ according to Eq. (\ref{U0}). 
It is easy to check that these embedded solitons are precisely the
ones reported in \cite{gromov0}. 
We have further determined the linear and nonlinear stabilities of these fixed points in 
equations (\ref{r}) and (\ref{k}), and found that they are all {\em stable}, both
linearly and nonlinearly. This means that these sech-shaped embedded solitons in
the perturbed CMKdV equation (\ref{uold}) are both linearly
and nonlinearly {\em stable}, 
despite the fact that such solitons reside inside the continuous spectrum of the
wave system and are in resonance with the continuous waves. 
They mark a distinct contrast with isolated embedded solitons 
and the continuous family of double-hump embedded solitons in
other physical systems, which were shown to be semi-stable 
\cite{yang_akylas,YangPRL99,YangStudies01,PeliYang}. 

The simplest way to demonstrate the linear and 
nonlinear stabilities of fixed points (\ref{fixpt})
is by drawing a phase portrait in the $(r, k)$ plane. 
Such a phase portrait with $\alpha=-0.1$ and $\gamma=0.1$ is shown in Fig. 1. 
At these parameter values, the fixed points are the $k=\frac{1}{2}$ line 
(dashed). Clearly, we see that these points are nonlinearly stable, 
and they attract a large region of initial parameters. 
At other $\alpha$ and $\gamma$ values where
$\alpha \gamma<0$, the phase portraits are qualitatively similar. 
If $\alpha \gamma >0$, 
the phase portrait is just the one of $\alpha \gamma <0$ with
$k$ replaced by $-k$. 

\begin{figure}
\begin{center}
\setlength{\epsfxsize}{4cm}\epsfbox{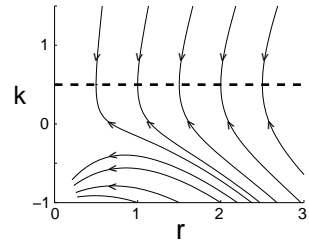}
\caption{Phase portrait of the dynamical system (\ref{r}) and (\ref{k})
for physical parameters $\alpha=-0.1$ and $\gamma=0.1$. }
\end{center}
\end{figure}

\vspace{-0.15cm}
Lastly, we confirm the above analytical results by direct numerical simulations
of Eq. (\ref{uold}). 
Our numerical scheme is the pseudo-spectral method along the $\xi$ direction, 
and the fourth-order Runge-Kutta method along the $z$ direction. 
Damping boundary conditions have also been used
to filter out the energy radiation emitted into the far field. 
A number of numerical simulations of Eq. (\ref{uold}), starting with
the soliton solution (\ref{u0}) at $z=0$, have been performed for various
small $\alpha$ and $\gamma$ values. Good agreement between the numerics
and the analysis has been observed. In addition, as
$\alpha$ and $\gamma$ approach zero, the difference between the
leading-order perturbation results [Eqs. (\ref{r}) and (\ref{k})]
and numerical values approaches zero, meaning 
that our perturbation theory is asymptotically accurate. 
For illustration purpose, 
we fix the physical parameters $\alpha=-0.1$ and $\gamma=0.1$,  
and choose two sets of initial amplitude and frequency values, 
$(r_0, k_0)$=(1.5, 1) and (1.5, 0), which
are on the opposite sides of the line of fixed points in Fig. 1. 
The simulation results are displayed in Fig. 2. 
For $(r_0, k_0)$=(1.5, 1), the pulse evolution is plotted in Fig. 2(a). 
Due to perturbations in Eq. (\ref{uold}), 
the speed of the pulse slowly increases. Thus the pulse turns around and eventually
approaches a steady positive speed. 
In Fig. 2(b), the field profiles at two distances are plotted. 
As predicted, a continuous-wave tail develops on the left of the embedded soliton. 
Note that the tail amplitude decreases with distance $z$, meaning
that the embedded soliton is stabilizing \cite{YangPRL99}. 
A direct comparison between 
the numerics and the leading-order perturbation theory in the phase
plane is shown in Fig. 2(d). 
We see that the numerical pulse frequency $k$ indeed approaches 
the theoretical value $\frac{1}{2}$. 

When $(r_0, k_0)$=(1.5, 0), the numerical pulse evolution is displayed in Fig. 2(c). 
In this case, the soliton slows down under perturbations, contrary
to the previous case. Examination of the radiation field indicates that
the continuous-wave tails emitted to the left of the soliton are also lessening, 
indicating that the embedded soliton is stabilizing again as the theory predicts. 
Comparison in Fig. 2(d) between the numerics and the theory for this case 
also shows good agreement. 

Physically, why are embedded solitons in Eq. (\ref{uold})
fully stable? There are two major reasons. First, these solitons
exist as a continuous family. In other words, in the neighborhood of each
embedded soliton, there are other embedded solitons nearby which have
higher or lower energy. Thus when perturbed, an embedded soliton
may always relax into an adjacent one, similar to the NLS-solitons. 
The second reason is that these embedded solitons are single-humped. 
Recall that double-humped embedded solitons in Eq. (\ref{psi}) also exist
as a continuous family, but they are not stable \cite{yang_akylas}. 
This is not surprising, as the instability of multi-hump solitons
is well documented in the literature. 
Embedded solitons in Eq. (\ref{uold}), however, are single-humped. 
Thus they could be fully stable as we have shown in this paper.

We note that for ultra-short pulses, $\beta =\mu$; while
in this paper, $|\alpha|, |\gamma| \ll 1$, i.e., $|\beta-6|\ll 1$ and $|\mu| \ll 1$. 
In the physical case $\beta =\mu$, a continuous family of single-hump
embedded solitons also exists \cite{gromov0}. In addition, 
Gromov, et al's numerical computations show that a sech pulse tends 
to one or a few single-hump embedded solitons \cite{gromov}. 
This numerical evidence, together with the above analytical results and
physical arguments, strongly suggests that embedded solitons in 
this physical case are also stable. 
A rigorous proof of this full stability in this physical case 
can be provided by an elaborate internal-perturbation method
\cite{PeliYang}, which we will pursue in the near future. 
It is also noted that for ultra-short pulses,
the Raman effect is non-negligible. 
But since embedded solitons without the Raman effect
are robust and stable, we can expect that the Raman effect on 
embedded solitons would be also a frequency downshift, 
similar to that on NLS solitons \cite{agrawal,hasegawa}. 

In summary, we have discovered fully-stable embedded solitons
in a physical model relevant for ultra-short pulse propagation in optical fibers. 
This finding dispels previous skepticism about the observability
of embedded solitons. We expect this work to have important 
implications to ultra-short pulse propagation. It 
would also stimulate further search for stable embedded solitons in other physical 
systems. 

This work was supported in part by a NASA 
grant. 

\begin{figure}
\setlength{\epsfxsize}{8cm}\epsfbox{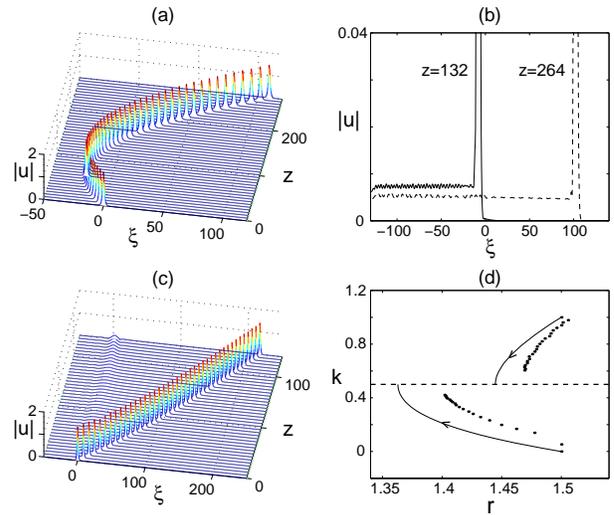}
\caption{
Numerical evolution of initial 
solitons (\ref{u0}) with $r_0=1.5$, $k_0=1$ [in (a, b)] and 0 [in (c)] under perturbations of
$\alpha=-0.1$ and $\gamma=0.1$ in Eq. (\ref{uold}). 
In (d), numerical and 
analytical results are compared in the
phase plane for both cases of $k_0=1$ and 0; 
solid lines: analytical; dots: numerical. 
}
\end{figure}

\end{multicols}

\end{document}